\documentclass{jpsj3}
\title{Momentum Dependent Local-Ansatz with Hybrid Wavefunction from Weak to Strong
Electron Correlations}

\author{
M. Atiqur R. \textsc{Patoary}\thanks{E-mail address:
k108609@eve.u-ryukyu.ac.jp} and  Yoshiro \textsc{Kakehashi}\thanks{yok@sci.u-ryukyu.ac.jp}
}

\inst{
Department of Physics and Earth Sciences,
Faculty of Science, \\
University of the Ryukyus, \\
1 Senbaru, Nishihara, Okinawa, 903-0213, Japan
}

\abst{
\begin{center}
( To be published in Journal of the Physical Society of Japan )
\end{center}

The variational theory of momentum dependent local-ansatz (MLA) has been 
generalized by introducing a hybrid (HB) wavefunction as a starting wavefunction,
whose potential can flexibly change from the Hartree-Fock type to the 
alloy-analogy  type by varying a weighting factor from zero to 
one. Numerical results based on the half-filled band Hubbard model on the 
hypercubic lattice in infinite dimensions show up that the new wavefunction 
yields the ground-state energy lower than that of  the Gutzwiller wavefunction (GW) in 
the whole Coulomb interaction regime. Calculated double occupation number 
is smaller than the result of the GW in the weak Coulomb interaction 
regime, and remains finite  in the strong regime. Furthermore, the momentum 
distribution shows a distinct momentum dependence, which is qualitatively 
different from that of the GW.}

\kword{ electron correlations, Hubbard model, variational method, Gutzwiller
wavefunction, local ansatz,  metal-insulator transition, critical
Coulomb interaction, infinite dimensions}

\begin{document}
\maketitle
\section{Introduction}
It is well recognized that electron correlations are 
essential for understanding the electronic structure, the magnetism, 
the metal-insulator transition, and the high-temperature superconductivity 
in solids~\cite{fulde95,kakeh12}. To describe  the correlations 
at the  ground-state,  various methods have been developed.
The variational approach is one of the simplest methods among 
them and has been applied to many systems as a practical 
tool~\cite{gutz63,gutz64,gutz65,stoll77,stoll78,stoll80,
dbaer87,mzie97,dbae00,bahe10,dtahara08,hyoka11,tmisawa11,tschi12}.  
A minimum basis set  is 
constructed  in the approach  by applying one-particle, 
two-particle, and  higher-order particle operators onto the 
Hartree-Fock (HF) wavefunction, and  their amplitudes are chosen to be best.

For the Hubbard-type Hamiltonian, the  Gutzwiller  wavefunction (GW) is one 
of useful wavefunctions, because of its simplicity and applicability 
to realistic systems~\cite{gutz63,gutz64,gutz65}.  When the intra-atomic 
Coulomb repulsion $U$ is large,  the double occupancy on the same orbital 
should be suppressed to avoid the energy loss due to the Coulomb repulsion 
$U$~\cite{hub-I, hub-II,hub-III}. The HF wavefunction does not describe such 
correlations because it consists of a single Slater determinant. 
Taking into account  these facts, Gutzwiller proposed a 
trial wavefunction which controls the probability amplitudes of doubly occupied 
states in the HF wavefunction by making use of a projection operator 
$\Pi_{i} (1-g\hat n_{i\uparrow}\hat n_{i\downarrow})$. 
Here $\hat n_{i\sigma}$ is the number operator for an electron on site $i$ 
with  spin $\sigma$, variational parameter $g$ reduces the amplitudes 
of doubly occupied states on  local orbitals.  Stollhoff and 
Fulde~\cite{stoll77,stoll78,stoll80} proposed a method called 
the local-ansatz approach (LA),  which is simpler than the GW in treatment. 
The LA wavefunction  takes into account the states created by local 
two-particle operators such  as the residual Coulomb interactions
$\{ O_{i} \}=\{ \delta \hat n_{i\uparrow}\delta \hat n_{i\downarrow} \}$. 
Here $\delta \hat n_{i\sigma}= 
\hat n_{i\sigma} - \langle \hat n_{i\sigma} \rangle_{\rm HF}$, $\langle \hat n_{i\sigma}
\rangle_{\rm HF}$ being the average electron number on 
site $i$ with spin $\sigma$ in the HF approximation.  

Though the GW and the LA are applicable for various correlated electron 
systems, they are not sufficient for the description of correlations from 
the weak to the strong interaction regimes. Indeed, the  Hilbert 
space expanded by the local operators  is not sufficient  to characterize  precisely 
the weakly correlated states; the LA does not reduce to the second-order 
perturbation theory in the weak correlation limit. The same  difficulty   
also arises for the  GW even in infinite dimensions. Moreover, in the strong 
Coulomb interaction regime, the GW yields the Brinkman-Rice atom ($i. e.$, 
no charge fluctuation on an atom) instead of the insulator solid in infinite 
dimensions~\cite{br70}. To overcome the difficulty 
in the weak Coulomb interaction regime and to improve the behaviors in 
the intermediate Coulomb interaction regime, we have recently proposed the 
momentum dependent local-ansatz wavefunction (MLA)~\cite{kakeh08,pat11,pat12},
and demonstrated that the MLA approach much improves both  the GW and 
the LA  in these regimes. In the MLA, we consider two-particle operators
in the momentum space with momentum dependent parameters and project them 
onto the local orbitals. With use of such local operators $\{\tilde O_i\}$, 
we construct the MLA wavefunction  as $| \Psi_{\rm MLA} \rangle = 
\prod_i (1-\tilde{O_i}) | \phi_{\rm HF} \rangle$. 
Here $| \phi_{\rm HF} \rangle$ is the HF wavefunction and $i$ denotes 
sites of atoms. The best local basis set is chosen by 
controlling the variational parameters in the momentum space.

Baeriswyl, on the other hand,  proposed a wavefunction called Baeriswyl 
wavefunction (BW) which accurately describes electron correlations in
the strong Coulomb interaction  regime~\cite{dbaer87, mzie97,dbae00,bahe10}. It  
is constructed by applying a hopping operator $\hat T$ onto the atomic  
wavefunction $|\Psi_{\infty}\rangle$; 
$|\Psi_{\rm BW}\rangle = e^{-\eta\hat T}|\Psi_{\infty}\rangle$. Here 
$\hat T=-\sum_{i,j,\sigma} t_{ij} {a}^\dag_{i\sigma} {a}_{j\sigma} $ is the
kinetic energy operator, $t_{ij}$ denotes the transfer integral between 
sites $i$ and $j$, $ a_{i\sigma}^{\dagger}$ ($ a_{i\sigma}$) being the creation 
(annihilation) operator for an electron on site $i$ with spin $\sigma$.  
The operator $e^{-\eta\hat T}$ with a variational parameter $\eta$ 
describes electron hopping from the atomic state and 
suppresses the configurations with high kinetic energy. 
The BW describes well the insulator state in the strong correlation 
regime. However, it is not easy to describe the metallic state from 
this viewpoint.

In order to  describe the correlations in the strong Coulomb interaction 
regime, we have recently proposed an improved MLA wavefunction~\cite{pat13}, which  starts
from the alloy-analogy (AA) wavefunction instead of the HF one.
The concept of the AA approximation  can be traced back to Hubbard's original
work on electron correlations~\cite{hub-III}. He considered that electrons move slowly from
site to site in the strong Coulomb interaction regime, so that an electron 
on a site with (without) opposite spin electron on the same site feels a 
potential $\epsilon_0+U$\, $(\epsilon_0)$, where  $\epsilon_0$ and $U$
denote the atomic level and the on-site Coulomb interaction parameter, respectively. 
The AA wavefunction is the ground-state wavefunction for the 
independent-particle Hamiltonian with such two kind of random potentials.
We found numerically that the MLA theory with the AA wavefunction describes 
the strongly  correlated regime reasonably, and  can go beyond the
GW in both  the  weak and the strong Coulomb interaction regimes. 

From the above discussions it is recognized that the MLA wavefunction 
can describe reasonably electron correlations from the weak to the intermediate 
Coulomb interaction regime and to the strong Coulomb interaction regime 
by choosing the starting wavefunction. In order to describe the whole Coulomb 
interaction regime on the same footing, we propose in this paper  a new
MLA wavefunction which starts from a hybrid (HB) wavefunction, 
and clarify the validity of our theory on the basis of the  
results of numerical  calculations for the half-filled  band Hubbard model.
The HB wavefunction is defined by the ground-state of  
the independent-particle Hamiltonian with a HB
potential consisting of the HF potential with a weight $1-w$ and the
AA potential with a weight $w$, and can vary  from the HF wavefunction
to the AA one via the new variational parameter $w$. 
Hereafter we call the new wavefunction the MLA-HB. 
We will demonstrate that the MLA-HB much improves both  
the GW and the  LA, and describes electron correlations from
the weak  to the strong Coulomb  interaction regime.

The outline of the paper is as follows. In the following section
we adopt the Hubbard model and introduce the  HB Hamiltonian as well as
the HB wavefunction. 
We will clarify the properties of the HB wavefunction, calculating 
the ground-state energy, the double occupation number and 
the momentum distribution in infinite dimensions. 
In \S 3, we present the correlated MLA-HB wavefunction
which starts from the HB wavefunction. We obtain  the 
ground-state energy within the single-site approximation (SSA), and 
derive the self-consistent equation for the momentum dependent 
variational parameters. We also obtain the double occupation number as well as the momentum 
distribution. In \S 4, we present  our results of  numerical 
calculations for the half-filled band Hubbard model on the hypercubic 
lattice in infinite dimensions. We discuss 
the ground-state energy, the double occupation number, the momentum  
distribution, and the quasiparticle weight as a function of 
the Coulomb interaction energy parameter, and  verify that
the present approach improves both the GW and the LA in the whole Coulomb 
interaction regime. We summarize our results in the last section and 
discuss the remaining problems.

\section{Hybrid wavefunction} 
We adopt in this paper the single-band Hubbard 
model\cite{hub-I,hub-II,hub-III} as follows. 
\begin{eqnarray}
H = \sum_{i \sigma} (\epsilon_{0}-\mu) \hat n_{i\sigma} 
+ \sum_{ij \sigma} t_{i j} \, a_{i \sigma}^{\dagger} a_{j \sigma} 
+ U \sum_{i} \, \hat n_{i \uparrow} \hat n_{i \downarrow} \ .
\label{hub}
\end{eqnarray}
Here $\epsilon_{0}$ ($\mu$) is the atomic level (chemical potential), 
$t_{ij}$ is the transfer integral between sites $i$ and $j$.  $U$
is the intra-atomic Coulomb energy parameter.  $ a_{i \sigma}^{\dagger}$
($ a_{i \sigma}$) denotes the creation (annihilation) operator for an
electron on site $i$ with spin $\sigma$, and 
$\hat n_{i\sigma}=a_{i\sigma}^{\dagger} a_{i \sigma}$ is the electron density
operator on site $i$ for spin $\sigma$.

In the HF approximation, we neglect the fluctuations 
$\delta \hat n_{i \uparrow} \delta \hat n_{i \downarrow}$  and replace 
the many-body Hamiltonian (\ref{hub}) with an effective Hamiltonian 
$H_{\rm HF}$ for  independent-particle system.
\begin{eqnarray}
H_{\rm HF}= \sum_{i\sigma}  (\epsilon_{0}-\mu+U\langle n_{i
-\sigma} \rangle_{\rm HF})\hat n_{i\sigma} 
+ \sum_{ij \sigma} t_{i j} \, a_{i \sigma}^{\dagger} a_{j \sigma}
- U \sum_{i} \, \langle n_{i \uparrow} \rangle_{\rm HF} 
\langle n_{i\downarrow} \rangle_{\rm HF} \ .
\label{hf}
\end{eqnarray}
Here $\langle \sim \rangle_{\rm HF}$ denotes the HF average 
$\langle \phi_{\rm HF}| (\sim) |\phi_{\rm HF}\rangle$, and $\langle n_{i
\sigma} \rangle_{\rm HF}$ is the average electron number on site $i$ with
spin $\sigma$. $|\phi_{\rm HF} \rangle$ denotes the ground-state 
wavefunction for the  HF Hamiltonian $H_{\rm HF}$.

In the AA approximation, we consider  the strong Coulomb interaction regime, 
where electrons with spin $\sigma$ move slowly from site to site due to 
electron correlations. Instead of the HF average potential 
$U\langle \hat n_{i-\sigma} \rangle_{\rm {HF}}$, electrons should feel 
there a potential $U\,(0)$, when the  opposite spin electron is 
occupied (unoccupied) on the same site. Hubbard regarded this system as 
an alloy with different random potentials $\epsilon_0+U$ and $\epsilon_0$. 
The AA Hamiltonian is then defined  by  
\begin{eqnarray}
H_{\rm AA} =\sum_{i\sigma}(\epsilon_0-\mu+U n_{i-\sigma})\hat n_{i\sigma}
+ \sum_{ij\sigma} t_{ij}  a_{i\sigma}^\dagger  a_{j\sigma} 
-U \sum_i (n_{i\uparrow}\langle n_{i \downarrow} \rangle_{\rm AA}
+ n_{i\downarrow}\langle n_{i \uparrow} \rangle_{\rm AA})\nonumber\\
+U \sum_i \langle n_{i \uparrow} \rangle_{\rm AA}
\langle n_{i \downarrow} \rangle_{\rm AA}.
\label{aa}
\end{eqnarray} 
Here  $\langle \sim \rangle_{\rm AA}$ denotes the AA average 
$\langle \phi_{\rm AA}| (\sim) |\phi_{\rm AA}\rangle$ with  
respect to the ground-state wavefunction $|\phi_{\rm AA} \rangle$
of the AA Hamiltonian $H_{\rm AA}$. Since  the electrons 
with opposite spin are  treated  to be static in the AA 
approximation, related operators $\{\hat n_{i-\sigma} \}$ are 
regarded as a random  static  $C$ number $n_{i-\sigma}$ ($0$ or $1$). 
Each configuration  $\{ n_{i\sigma} \}$ is considered as a 
snapshot in time development.

The HF Hamiltonian works best in the weakly correlated regime, while 
the AA Hamiltonian works better in the strongly correlated regime. 
In order to obtain a good starting wavefunction for any interaction 
strength $U$, we introduce the HB Hamiltonian which is 
a linear combination of both the HF and the AA Hamiltonian as follows.
\begin{eqnarray}
H_{\rm HB}& = & \sum_{i\sigma}  (\epsilon_{0}-\mu
+\overline U\langle n_{i-\sigma} \rangle_{0}
+\widetilde U n_{i-\sigma}) \hat n_{i\sigma} 
+ \sum_{ij \sigma} t_{i j} \, a_{i \sigma}^{\dagger} a_{j \sigma}\nonumber\\
& & \hspace*{5mm} 
- (\overline U- \widetilde U) \sum_{i} \, 
\langle n_{i \uparrow} \rangle_{0} \langle n_{i\downarrow} \rangle_{0}
-\widetilde U \sum_i (n_{i\uparrow}\langle n_{i \downarrow} \rangle_{0}
+ n_{i\downarrow}\langle n_{i \uparrow} \rangle_{0}) \ .
\label{hb}
\end{eqnarray}
Here $\langle \sim \rangle_{0}$ denotes the HB average 
$\langle \phi_{0}| (\sim) |\phi_{0}\rangle$ with respect to 
the ground-state $|\phi_{0}\rangle$ of the HB Hamiltonian,
$\overline U=(1-w)U$ and $\widetilde U=w U$. We introduced a variational 
parameter $w$. Note that $H_{\rm HB}$ reduces to the  HF 
Hamiltonian when $w=0$,  while $H_{\rm HB}$  reduces to the AA when $w=1.0$.

The ground-state energy $E$ satisfies the following inequality for a normalized 
wavefunction $|\phi_{0}\rangle$.
\begin{eqnarray}
E \leq \langle \phi_{0} |H| \phi_{0} \rangle =\langle H_{\rm HB} \rangle_{0}\ .
\end{eqnarray}
The HB ground-state energy per atom is  obtained 
by taking the configurational average.
\begin{eqnarray}
\overline{ \langle H \rangle}_{\rm{HB}}= n\mu + 2\int^{0}_{-\infty}  
\epsilon \, \overline{\rho_{i\sigma}(\epsilon)} \, d\epsilon 
- (\overline U- \widetilde U) \overline{\langle n_{i\uparrow} 
\rangle_{0} \langle n_{i\downarrow} \rangle}_0
-\widetilde U  (\overline{n_{i\uparrow}\langle n_{i\downarrow} \rangle}_{0}
+ \overline{n_{i\downarrow}\langle n_{i\uparrow} \rangle}_0)\ .
\label{ehb}
\end{eqnarray}
Here we assumed the system with one atom per unit cell.
$\langle H \rangle_{\rm HB}$ denotes the HB average 
$\langle \phi_{0} |H| \phi_{0} \rangle$. The upper bar denotes the configurational 
average and $n$ is the electron number per atom. $\rho_{i\sigma}(\epsilon)$ is
the local density of states (DOS) and is obtained from the one-electron Green function. 
\begin{eqnarray}
\rho_{i\sigma}(\epsilon)= -\dfrac{1}{\pi}\, \rm {Im} \, \it G_{ii\sigma}(z) \ .
\end{eqnarray}
The Green function $G_{ii\sigma}(z)$ is defined by 
\begin{eqnarray}
 G_{ii\sigma}(z)= [(z-\mathbf{H}_\sigma)^{-1} ]_{ii}\ .
\end{eqnarray}
Note that $z=\epsilon+i\delta$, $\delta$ being the infinitesimal 
positive number. $(\mathbf{H_\sigma})_{ij}$ is the one-electron Hamiltonian 
matrix  for the HB Hamiltonian (\ref{hb}), which is defined by 
\begin{eqnarray}
(\mathbf{H_\sigma})_{ij}=(\epsilon_{0}-\mu
+\overline U\langle n_{i-\sigma} \rangle_{0}
+\widetilde U n_{i-\sigma}) \delta_{ij}
+ t_{i j} (1-\delta_{ij})\ .
\label{hbm}
\end{eqnarray} 
The average electron number $\langle n_{i\sigma}\rangle_0$ with respect to the 
HB Hamiltonian (\ref{hb}) is given as
\begin{eqnarray}
\langle n_{i\sigma}\rangle_0=\int f(\epsilon) \rho_{i\sigma}(\epsilon) \, d\epsilon  \ ,
\end{eqnarray} 
$f(\epsilon)$ being the Fermi distribution function.
 
To obtain the local  DOS, we make use of the coherent potential
approximation (CPA)~\cite{shiba71,ehr76}. In the CPA, we replace the random potentials at the 
surrounding sites  with a coherent potentials $\Sigma_{\sigma}(z)$.
The on-site impurity Green function $G_{ii\sigma}(z)$
is then obtained as follows.
\begin{eqnarray}
G_{ii\sigma}(z)=\frac{1}{F_{\sigma}(z)^{-1}-\epsilon_0
+\mu-\overline{U}\langle n_{i-\sigma}\rangle_0-\widetilde U n_{i-\sigma}+\Sigma_\sigma(z)} \ .
\end{eqnarray}
Here $F_{\sigma}(z)$ is the on-site Green function for the coherent system in which 
all the random potentials have been replaced by the coherent ones. It is given by
\begin{eqnarray}
F_{\sigma}(z)= 
\int \frac{\rho(\epsilon) \, d\epsilon}
{z - \Sigma_{\sigma}(z)-\epsilon} \ .
\label{cohg2}
\end{eqnarray}
Here $\rho(\epsilon)$ is the DOS per site for the noninteracting system.
  
The coherent potential $\Sigma_{\sigma}(z)$ is determined from a
self-consistent condition.
\begin{eqnarray}
\overline{G_{00\sigma}(z)} = F_\sigma(z) \ .
\label{cpa}
\end{eqnarray}
The configurational average of  the impurity Green function is now given as  
\begin{eqnarray}
\overline{G_{00\sigma}(z)}=\sum_{\alpha} P_\alpha G^{\alpha}_{00\sigma}(z) \ .
\label{gf}
\end{eqnarray}
Here  $\alpha= \, 00, \, 10, \, 01, \, 11 $ denotes the on-site electron 
configuration ($n_{0\uparrow}$,$ n_{0\downarrow}$). 
Alternative notation 
$\nu=0$ (empty on a site), $1\uparrow$ (occupied by an electron with spin $\uparrow$ ),
$1\downarrow$ (occupied by an electron with spin $\downarrow$ ) and 
$2$ (occupied by 2 electrons) is also useful. 
In this case, the probability $P_\alpha$ for the configuration $\alpha$
is expressed as $P_0$, $P_{1\uparrow}$, 
$P_{1\downarrow}$  and $P_2$. 

The impurity Green functions in eq. (\ref{gf}) for each configuration  are 
given as follows.
\begin{eqnarray}
{G^{00}_{00\sigma}(z)}=\frac{1}{F_{\sigma}(z)^{-1}-\epsilon_0+\mu-\overline{U}\langle n_{-\sigma}\rangle_{00}+\Sigma_\sigma(z)}\ ,
\end{eqnarray}
\begin{eqnarray}
{G^{10}_{00\uparrow}(z)}=\frac{1}{F_{\sigma}(z)^{-1}-\epsilon_0+\mu-\overline{U}\langle n_{\downarrow}\rangle_{10}+\Sigma_\sigma(z)} \ ,
\end{eqnarray}
\begin{eqnarray}
{G^{10}_{00\downarrow}(z)}=\frac{1}{F_{\sigma}(z)^{-1}-\epsilon_0+\mu-\overline{U}\langle n_{\uparrow}\rangle_{10}-\widetilde U+\Sigma_\sigma(z)} \ ,
\end{eqnarray}
\begin{eqnarray}
{G^{01}_{00\uparrow}(z)}=\frac{1}{F_{\sigma}(z)^{-1}-\epsilon_0+\mu-\overline{U}\langle n_{\downarrow}\rangle_{01}-\widetilde U+\Sigma_\sigma(z)} \ ,
\end{eqnarray}
\begin{eqnarray}
{G^{01}_{00\downarrow}(z)}=\frac{1}{F_{\sigma}(z)^{-1}-\epsilon_0+\mu-\overline{U}\langle n_{\uparrow}\rangle_{01}+\Sigma_\sigma(z)} \ ,
\end{eqnarray}
\begin{eqnarray}
{G^{11}_{00\sigma}(z)}=\frac{1}{F_{\sigma}(z)^{-1}-\epsilon_0+\mu-\overline{U}\langle n_{-\sigma}\rangle_{11}-\widetilde U+\Sigma_\sigma(z)} \ .
\end{eqnarray}
Here the electron number in the denominator is given by
\begin{eqnarray}
\langle n_{\sigma}\rangle_{\alpha}= \int  f (\epsilon) \rho^{\alpha}_{\sigma}(\epsilon)\, d \epsilon  \ ,
\end{eqnarray}
\begin{eqnarray}
\rho^{\alpha}_{\sigma}(\epsilon)= -\dfrac{1}{\pi}\, \rm {Im} \, G^{\alpha}_{00\sigma}(z) \ .
\label{doshb}
\end{eqnarray}
Furthermore, the average DOS in the second term at the right hand side (rhs) of eq. (\ref{ehb})
is given by 
\begin{eqnarray}
\overline{\rho_{i\sigma}(\epsilon)}= -\dfrac{1}{\pi}\, \rm {Im} \, \overline{G_{00\sigma}(z)} \ .
\end{eqnarray}

It should be noted that $P_0+ P_{1\uparrow}+P_{1\downarrow}+P_2=1$,  and 
the probability of finding an electron with spin $\uparrow(\downarrow)$ on 
a site is given by $P_{\uparrow(\downarrow)}=P_{1\uparrow(1\downarrow)}+P_2$. 
Three statistical probabilities $P_0, \, P_{1\uparrow}, \, \rm {and} 
\,P_{1\downarrow} $  therefore depend on the probability $P_2$. The expression 
of $P_2$ is given as follows as shown in  Appendix A.
\begin{eqnarray}
P_2=\dfrac{(1-w)\langle n_{\uparrow}\rangle^{2}_{00}+(P_\uparrow+P_\downarrow)
\{1/2\, w \,\langle n_{\uparrow}\rangle_{01} + (1-w)\langle n_{\uparrow}\rangle_{10}
\langle n_{\uparrow}\rangle_{01}- \langle n_{\uparrow}\rangle^{2}_{00}\}}
{1-w(\langle n_{\uparrow}\rangle_{11}-\langle n_{\uparrow}\rangle_{01})
-(1-w)(\langle n_{\uparrow}\rangle^{2}_{00}-2\langle n_{\uparrow}\rangle_{10}
\langle n_{\uparrow}\rangle_{01}+\langle n_{\uparrow}\rangle^{2}_{11} )} \, . 
\label{p2}
\end{eqnarray} 

The double occupation numbers at the rhs of eq. (\ref{ehb}) are obtained in the 
SSA as follows.
\begin{eqnarray}
\overline{\langle n_{i \uparrow} n_{i\downarrow} \rangle}_0=
\overline{\langle n_{i \uparrow} \rangle_{0} \langle n_{i\downarrow} \rangle}_0=\sum_{\alpha} 
P_\alpha \langle n_{ \uparrow} \rangle_{\alpha} \langle n_{\downarrow}\rangle_{\alpha} \ ,
\label{dblhb}
\end{eqnarray} 
\begin{eqnarray}
\overline{n_{i\uparrow}\langle n_{ i\downarrow} \rangle}_{0}+ \overline{n_{i\downarrow}\langle n_{i\uparrow} \rangle}_0
 \, = \, (P_\uparrow+P_\downarrow)\langle n_{ \uparrow} \rangle_{01}
 + 2 P_2 \, (\langle n_{ \uparrow} \rangle_{11}-\langle n_{ \uparrow} \rangle_{01}) \ . 
\end{eqnarray} 
The  momentum distribution in  the HB scheme is given by 
\begin{eqnarray}
\overline{\langle n_{k\sigma} \rangle}_{0}= \int 
f (\epsilon) {\rho_{k\sigma}(\epsilon)}\, d \epsilon   \ ,
\label{nkhb}
\end{eqnarray}
\begin{eqnarray}
{\rho_{k\sigma}(\epsilon)} =  -\dfrac{1}{\pi}\, \rm {Im} \,\it F_{k\sigma} \ ,
\label{mdos}
\end{eqnarray}
\begin{eqnarray}
{F}_{k\sigma} = 
\frac{1}{z - \Sigma_{\sigma}(z)-\epsilon_k} \ .
\end{eqnarray}
Here $\epsilon_k$ is the eigen value of $t_{ij}$ with momentum $k$.
%========================================================================
\section{Local-Ansatz ${+}$ Hybrid Wavefunction Approach  with 
Momentum Dependent Variational Parameters} 
The momentum dependent local-ansatz (MLA) wavefunction is based on the 
local-ansatz (LA) proposed 
by Stollhoff and Fulde: $|\Psi_{\rm LA}\rangle = \big[ \prod_{i} 
(1 - \eta^{}_{\rm \, LA} O_{i}) \big]|\phi_{\rm HF} 
\rangle$~\cite{stoll77,stoll78,stoll80}. Here 
$ O_{i}=\delta n_{i\uparrow}\delta n_{i\downarrow}$ are the 
residual interaction, the  amplitude $\eta^{}_{\rm \, LA}$ is determined 
variationally. The operators $\{O_i \}$ expand the Hilbert space 
to describe the weak Coulomb interaction regime.
The LA however does not yield the exact result in the weak interaction limit.
The MLA wavefunction is constructed to describe exactly the weak limit 
as follows~\cite{kakeh08,pat11}.
\begin {equation}
 | \Psi_{\rm MLA} \rangle = \prod_i (1-\tilde{O_i}) | \phi_{\rm HF} \rangle \ ,
\label{wmlahf}
\end{equation} 
\begin{eqnarray}
\tilde{O_i} = \sum_{k_1 k'_1 k_2 k'_2} 
\langle k'_1 | i\rangle \langle i | k_1 \rangle 
\langle k'_2 | i\rangle \langle i | k_2 \rangle 
\ \eta_{k'_2 k_2 k'_1 k_1} 
\delta ({a_{k'_2 \downarrow}^\dagger} a_{k_2 \downarrow }) 
\delta ({a_{k'_1 \uparrow }^\dagger} a_{k_1 \uparrow }) \ . 
\end{eqnarray} 
Here $\langle i|k \rangle 
= \exp (-i\boldsymbol{k}\cdot \boldsymbol{R}_{i}) / \sqrt{N}$ 
is an overlap integral between the localized orbital and the Bloch state with
momentum $\boldsymbol{k}$, $\boldsymbol{R}_{i}$ denotes atomic
position, and $N$ is the number of sites. $ \eta_{k'_2 k_2 k'_1 k_1} $
is a momentum dependent variational parameter. 
$ a_{k \sigma}^{\dagger}$ ($ a_{k \sigma}$) denotes a creation 
(annihilation) operator for an electron with momentum $\boldsymbol{k}$ 
and spin $\sigma$, and 
$\delta(a^{\dagger}_{k^{\prime}\sigma}a_{k\sigma})= 
a^{\dagger}_{k^{\prime}\sigma}a_{k\sigma} - \langle
a^{\dagger}_{k^{\prime}\sigma}a_{k\sigma} \rangle_{\rm HF}$.
Note that the local operator $\tilde{O_i}$ reduces to $\eta_{\rm LA}{O_i}$
when $ \eta_{k'_2 k_2 k'_1 k_1} \rightarrow \eta_{\rm LA}$.
The best wavefunction is chosen by minimizing the energy with respect to 
the variational parameters in the momentum space.

In this work, we generalize the wavefunction (\ref{wmlahf}) to be  suitable 
in both the strong and the weak Coulomb interaction regime; we adopt the 
HB ground-state  wavefunction $|\phi_{0}\rangle$ for the Hamiltonian $H_{\rm HB}$
(\ref{hb}),  and apply a new correlator $\prod_i (1-\tilde{O_i})$ as follows. 
\begin {equation}
 | \Psi_{\rm MLA-HB} \rangle = \prod_i (1-\tilde{O_i}) | \phi_{0} \rangle .
\label {wmlahb} 
\end{equation}  
Note that the local operators $\{\tilde O_i\}$  have been modified as 
follows. 
\begin{eqnarray}
\tilde{O_i} = \sum_{\kappa'_2 \kappa_2 \kappa'_1 \kappa_1} 
\langle \kappa'_1 | i\rangle \langle i | \kappa_1 \rangle 
\langle \kappa'_2 | i\rangle \langle i | \kappa_2 \rangle 
\ \eta_{\kappa'_2 \kappa_2 \kappa'_1 \kappa_1} 
\delta ({a_{\kappa'_2 \downarrow}^\dagger} a_{\kappa_2 \downarrow }) 
\delta ({a_{\kappa'_1 \uparrow }^\dagger} a_{\kappa_1 \uparrow }) \ . 
\end{eqnarray} 
Here $\eta_{\kappa'_2 \kappa_2 \kappa'_1 \kappa_1}$ is a variational parameter,
$ a_{\kappa \sigma}^{\dagger}$ and $ a_{\kappa \sigma}$ are the creation 
and annihilation operators which diagonalize the Hamiltonian $H_{\rm HB}$ 
(\ref{hb}), and 
$\delta(a^{\dagger}_{\kappa^{\prime}\sigma}a_{\kappa\sigma})= 
a^{\dagger}_{\kappa^{\prime}\sigma}a_{\kappa\sigma} - \langle
a^{\dagger}_{\kappa^{\prime}\sigma}a_{\kappa\sigma} \rangle_{0}$.
It should be noted that the MLA-HB wavefunction (\ref{wmlahb}) reduces to the MLA-HF 
with the uniform potential $U \langle n_{i-\sigma}\rangle_{\rm HF}$ 
when the  variational parameter $w=0$, and  reduces to the MLA-AA
with the random potential $U n_{i-\sigma}$  when $w=1$. The MLA-HB 
wavefunction  interpolates between the two wavefunctions. 

The ground-state energy $E$ satisfies the following inequality for any 
wavefunction $|\Psi\rangle$.
\begin{eqnarray}
E & \leq & \dfrac{\langle \Psi |H| \Psi \rangle}{\langle \Psi | \Psi \rangle} 
= \langle H \rangle_{\rm HB}+ N\epsilon_c \ .
\end{eqnarray}
Here $\langle H \rangle_{\rm HB}$ denotes the energy 
for the HB wavefunction. $\epsilon_c$ is the correlation energy per atom defined by
\begin{eqnarray}
N\epsilon_c =\dfrac{\langle \Psi |\widetilde {H}| \Psi \rangle}
{ \langle \Psi | \Psi \rangle} \ ,
\end{eqnarray}
with $\widetilde{H}=H - \langle H \rangle_{\rm HB}$.  Since it depends 
on the electron configuration $\{ n_{i\sigma}\}$ via the AA potential,
we have to take into account the configurational average at the
end. To determine the variational parameters, we minimize the 
ground-state energy.

It is not easy to calculate exactly the correlation energy with use of 
the HB wavefunction (\ref{wmlahb}). Therefore, we adopt here 
the single-site approximation (SSA). The average of $\langle \tilde{A} \rangle$  of an operator 
$\tilde{A}= A-\langle A \rangle_0$  with respect to the  
wavefunction (\ref{wmlahb}) is then given as follows.
\begin{eqnarray}
\langle \tilde{A} \, \rangle = 
\sum_{i} \dfrac{\langle (1 - \tilde{O}^{\dagger}_{i}) \tilde{A}
(1 - \tilde{O}_{i}) \rangle_{0}}
{\langle (1 - \tilde{O}^{\dagger}_{i})(1 - \tilde{O}_{i}) \rangle_{0}} \ .
\label{ava}
\end{eqnarray}
The detailed derivation of the above formula has been given in Appendix 
A of our paper~\cite{kakeh08}. Making use of the above formula,
the correlation energy per atom is obtained as follows. 
\begin{eqnarray}
\epsilon_{\rm c} = \dfrac{-\langle
 \tilde{O}^{\dagger}_{i}\tilde{H}\rangle_{0} -
\langle \tilde{H} \tilde{O}_{i} \rangle_{0} + 
\langle \tilde{O}^{\dagger}_{i}\tilde{H}\tilde{O}_{i}\rangle_{0}}
{1 + \langle \tilde{O}^{\dagger}_{i}\tilde{O}_{i} \rangle_{0}} \ .
\label{ec}
\end{eqnarray}

Each term in the correlation energy (\ref{ec}) can be calculated by
making use of Wick's theorem as follows.
\begin{eqnarray}
\langle \tilde{H} \tilde{O}_{i} \rangle_{0} & = &  U 
\sum_{\kappa'_2 \kappa_2 \kappa'_1 \kappa_1}
\langle \kappa'_1 | i\rangle \langle i | \kappa_1 \rangle
\langle \kappa'_2 | i\rangle \langle i | \kappa_2 \rangle
\sum_{j}
\langle \kappa_1|j \rangle \langle j|\kappa'_1 \rangle 
\langle \kappa_2|j \rangle \langle j| \kappa'_2\rangle \nonumber \\
& & \times
\eta_{\kappa'_2 \kappa_2 \kappa'_1 \kappa_1}\tilde f_{\kappa'_2 \kappa_2 \kappa'_1 \kappa_1}\ ,
\label{ho}
\end{eqnarray}
\begin{eqnarray}
\langle \tilde{O}^{\dagger}_{i}\tilde{H} \rangle_{0} =
\langle \tilde{H} \tilde{O}_{i} \rangle^{\ast}_{0} \ ,
\hspace{88mm}
\label{oh}
\end{eqnarray}
\begin{eqnarray}
\langle \tilde{O}^{\dagger}_{i}\tilde{H}\tilde{O}_{i}\rangle_{0} & = & 
\sum_{\kappa'_2 \kappa_2 \kappa'_1 \kappa_1}
\langle i | \kappa'_1 \rangle\langle \kappa_1 | i\rangle 
\langle i | \kappa'_2 \rangle\langle \kappa_2 | i\rangle  \,
\eta^{\ast}_{\kappa'_2 \kappa_2 \kappa'_1 \kappa_1}  
\tilde f_{\kappa'_2 \kappa_2 \kappa'_1 \kappa_1} \nonumber \\
& & \times
\sum_{\kappa'_4 \kappa_4 \kappa'_3 \kappa_3} 
 \langle \kappa'_3 | i\rangle \langle i | \kappa_3 \rangle
\langle \kappa'_4 | i\rangle \langle i | \kappa_4 \rangle 
\big [ \Delta E_{\kappa'_2 \kappa_2 \kappa'_1 \kappa_1}
\delta_{\kappa_1 \kappa_3}\delta_{\kappa'_1\kappa'_3}
\delta_{\kappa_2\kappa_4}\delta_{\kappa'_2\kappa'_4} \nonumber \\
 & & 
 + U_{\kappa'_2 \kappa_2 \kappa'_1 \kappa_1
\kappa'_4 \kappa_4 \kappa'_3 \kappa_3}\big]
\eta_{\kappa'_4 \kappa_4 \kappa'_3 \kappa_3} \ ,
\label{oho}
\end{eqnarray}
\begin{eqnarray}
U_{\kappa'_2 \kappa_2 \kappa'_1 \kappa_1
\kappa'_4 \kappa_4 \kappa'_3 \kappa_3} & = & 
U \sum_{j} [
\langle j|\kappa_1 \rangle \langle \kappa_3|j \rangle 
f(\tilde{\epsilon}_{\kappa_3\uparrow})\delta_{\kappa'_{1}\kappa'_{3}}
- \langle \kappa'_{1}|j \rangle \langle j|\kappa'_{3} \rangle
(1 - f(\tilde{\epsilon}_{\kappa'_{3}\uparrow})) \delta_{\kappa_{1}\kappa_{3}}
]   \nonumber \\ 
& & \hspace {-3mm}
\times 
[\langle j|\kappa_{2} \rangle \langle \kappa_{4}|j \rangle 
f(\tilde{\epsilon}_{\kappa_{4}\downarrow})\delta_{\kappa'_{2}\kappa'_{4}}
- \langle \kappa'_{2}|j \rangle \langle j|\kappa'_{4} \rangle
(1 - f(\tilde{\epsilon}_{\kappa'_{4}\downarrow})) \delta_{\kappa_{2}\kappa_{4}}
] \ ,
\label{uk}
\end{eqnarray}
\begin{eqnarray}
\langle \tilde{O}^{\dagger}_{i}\tilde{O}_{i} \rangle_{0} & = &
\sum_{\kappa'_2 \kappa_2 \kappa'_1 \kappa_1}
|\langle \kappa'_1 | i\rangle|^{2} |\langle \kappa_1 |i  \rangle|^{2}
|\langle \kappa'_2 | i\rangle|^{2} |\langle \kappa_2 |i  \rangle|^{2}\,
|\eta_{\kappa'_2 \kappa_2 \kappa'_1 \kappa_1}|^{2} \,
\tilde f_{\kappa'_2 \kappa_2 \kappa'_1 \kappa_1} \ . 
\hspace{00mm}
\label{oo}
\end{eqnarray}
Here $\tilde f_{\kappa'_2 \kappa_2 \kappa'_1 \kappa_1}$ is a fermi factor
of two-particle excitations which is defined by 
$\tilde f_{\kappa'_2 \kappa_2 \kappa'_1 \kappa_1}
= f(\tilde{\epsilon}_{\kappa_1\uparrow})
(1-f(\tilde{\epsilon}_{\kappa'_{1}\uparrow}))
f(\tilde{\epsilon}_{\kappa_2\downarrow})
(1-f(\tilde{\epsilon}_{\kappa'_2\downarrow}))$, $f(\epsilon)$ is the Fermi distribution
function at zero temperature, $\tilde{\epsilon}_{\kappa\sigma}=\epsilon_{\kappa\sigma}-\mu$,
and  $\epsilon_{\kappa\sigma}$ is the one-electron energy
eigen value for the HB Hamiltonian.
Moreover,  
 $ \Delta E_{\kappa'_2 \kappa_2 \kappa'_1 \kappa_1} = 
\epsilon_{\kappa'_{2}\downarrow} - \epsilon_{\kappa_{2}\downarrow}
+ \epsilon_{\kappa'_{1}\uparrow} - \epsilon_{\kappa_{1}\uparrow}$
is a two-particle excitation energy. 

The above expressions (\ref{ho}) and (\ref{uk}) contain nonlocal terms 
via summation over $j$ ({\it i.e.}, $\sum_{j}$).
We thus make additional SSA that we only take into account the local term ($j=i$),
so that $\langle \tilde{H} \tilde{O}_{i}\rangle_{0} 
(= \langle \tilde{O}^{\dagger}_{i}\tilde{H} \rangle_{0}^{\ast})$
and  $\langle \tilde{O}^{\dagger}_{i}\tilde{H}\tilde{O}_{i}\rangle_{0}$ reduce as follows. 
\begin{eqnarray}
\langle \tilde{H} \tilde{O}_{i} \rangle_{0}  = U  \sum_{\kappa'_2 \kappa_2 \kappa'_1 \kappa_1}
|\langle \kappa'_1 | i\rangle|^{2} |\langle \kappa_1 |i  \rangle|^{2}
|\langle \kappa'_2 | i\rangle|^{2} |\langle \kappa_2 |i  \rangle|^{2}\,
\eta_{\kappa'_2 \kappa_2 \kappa'_1 \kappa_1} \,
\tilde f_{\kappa'_2 \kappa_2 \kappa'_1 \kappa_1}  \ ,
\label{hor0}
\end{eqnarray}
\begin{eqnarray}
\langle \tilde{O}^{\dagger}_{i}\tilde{H}\tilde{O}_{i}\rangle_{0} & = & 
\sum_{\kappa'_2 \kappa_2 \kappa'_1 \kappa_1}
|\langle \kappa'_1 | i\rangle|^{2} |\langle \kappa_1 |i  \rangle|^{2}
|\langle \kappa'_2 | i\rangle|^{2} |\langle \kappa_2 |i  \rangle|^{2}\,\nonumber \\
& & \hspace*{00mm}
\times
\eta^{\ast}_{\kappa'_2 \kappa_2 \kappa'_1 \kappa_1} \,
\tilde f_{\kappa'_2 \kappa_2 \kappa'_1 \kappa_1} 
\bigg[ 
\Delta E_{\kappa'_{2}\kappa_{2}\kappa'_{1}\kappa_{1}} 
\eta_{\kappa'_2 \kappa_2 \kappa'_1 \kappa_1} \nonumber \\
& & \hspace*{00mm}
+ U \Big\{
\sum_{\kappa_{3}\kappa_{4}}|\langle \kappa_3 | i\rangle|^{2} |\langle \kappa_4 | i \rangle|^{2}
f(\tilde{\epsilon}_{\kappa_{3}\uparrow})f(\tilde{\epsilon}_{\kappa_{4}\downarrow}) \,
\eta_{\kappa'_2 \kappa_4 \kappa'_1 \kappa_3}\nonumber \\
& &\hspace*{00mm}
- \sum_{\kappa'_{3}\kappa_{4}}|\langle \kappa'_3 | i\rangle|^{2} |\langle \kappa_4 | i \rangle|^{2}
(1-f(\tilde{\epsilon}_{\kappa'_{3}\uparrow}))f(\tilde{\epsilon}_{\kappa_{4}\downarrow})\,
\eta_{\kappa'_2 \kappa_4 \kappa'_3 \kappa_1}  \nonumber \\
& & \hspace*{00mm}
- \sum_{\kappa_{3}\kappa'_{4}}|\langle \kappa_3 | i\rangle|^{2} |\langle \kappa'_4 | i \rangle|^{2}
 f(\tilde{\epsilon}_{\kappa_{3}\uparrow})(1-f(\tilde{\epsilon}_{\kappa'_{4}\downarrow}))\,
\eta_{\kappa'_4 \kappa_2 \kappa'_1 \kappa_3}\nonumber \\
& &\hspace*{00mm}
+ \sum_{\kappa'_{3}\kappa'_{4}}|\langle \kappa'_3 | i\rangle|^{2} |\langle \kappa'_4 | i \rangle|^{2}
 (1-f(\tilde{\epsilon}_{\kappa'_{3}\uparrow}))(1-f(\tilde{\epsilon}_{\kappa'_{4}\downarrow}))\,
\eta_{\kappa'_4 \kappa_2 \kappa'_3 \kappa_1}
\Big\}
\bigg] \ .
\label{ohor0}
\end{eqnarray}

In order to obtain the variational parameters 
$\{ \eta_{\kappa'_2 \kappa_2 \kappa'_1 \kappa_1} \}$, we minimize the 
correlation energy $\epsilon_{\rm c}$, {\it i.e.}, eq. (\ref{ec}) with 
eqs. (\ref{oo}), (\ref{hor0}), and (\ref{ohor0}).
The self-consistent equations for 
$\{\eta_{\kappa'_2 \kappa_2 \kappa'_1 \kappa_1} \}$
in the SSA are given as follows.
\begin{eqnarray}
(\Delta E_{\kappa'_2 \kappa_2 \kappa'_1 \kappa_1} - \epsilon_{\rm c})
\eta_{\kappa'_2 \kappa_2 \kappa'_1 \kappa_1}  
&+& U\bigg[
\sum_{\kappa_{3}\kappa_{4}}|\langle \kappa_3 | i\rangle|^{2} |\langle \kappa_4 | i \rangle|^{2}
f(\tilde{\epsilon}_{\kappa_{3}\uparrow})f(\tilde{\epsilon}_{\kappa_{4}\downarrow}) \,
\eta_{\kappa'_2 \kappa_4 \kappa'_1 \kappa_3}\hspace*{3mm}\nonumber \\
& &\hspace*{-30mm}
- \sum_{\kappa'_{3}\kappa_{4}}|\langle \kappa'_3 | i\rangle|^{2} |\langle \kappa_4 | i \rangle|^{2}
(1-f(\tilde{\epsilon}_{\kappa'_{3}\uparrow}))f(\tilde{\epsilon}_{\kappa_{4}\downarrow})\,
\eta_{\kappa'_2 \kappa_4 \kappa'_3 \kappa_1}  \nonumber \\
& & \hspace*{-30mm}
- \sum_{\kappa_{3}\kappa'_{4}}|\langle \kappa_3 | i\rangle|^{2} |\langle \kappa'_4 | i \rangle|^{2}
 f(\tilde{\epsilon}_{\kappa_{3}\uparrow})(1-f(\tilde{\epsilon}_{\kappa'_{4}\downarrow}))\,
\eta_{\kappa'_4 \kappa_2 \kappa'_1 \kappa_3}\nonumber \\
& &\hspace*{-30mm}
+ \sum_{\kappa'_{3}\kappa'_{4}}|\langle \kappa'_3 | i\rangle|^{2} |\langle \kappa'_4 | i \rangle|^{2}
 (1-f(\tilde{\epsilon}_{\kappa'_{3}\uparrow}))(1-f(\tilde{\epsilon}_{\kappa'_{4}\downarrow}))\,
\eta_{\kappa'_4 \kappa_2 \kappa'_3 \kappa_1}
\Big] = U \, . 
\label{eqeta}
\end{eqnarray}
It is not easy to find the solution of eq. (\ref{eqeta}) for the
intermediate strength of Coulomb interaction $U$. To solve  the equation 
approximately, we make use of an interpolate solution which  is valid 
in both the weak  Coulomb interaction limit and the atomic limit. 
Note that the first term at the left hand side (lhs) of eq. (\ref{eqeta}) 
is dominant  and  the second term is negligible in the weak Coulomb 
interaction limit. In the atomic limit, the momentum dependence  of
$\eta_{\kappa'_2 \kappa_2 \kappa'_1 \kappa_1}$ is negligible. Thus, 
we approximate $\{ \eta_{\kappa^{\prime}_{2}\kappa_{2}\kappa^{\prime}_{1}\kappa_{1}} \}$ 
in the second term at the lhs of eq. (\ref{eqeta}) with a 
momentum independent parameter $\eta$ which is suitable for the
atomic region. Solving the equation, we obtain
\begin{eqnarray}
\eta_{\kappa^{\prime}_{2}\kappa_{2}\kappa^{\prime}_{1}\kappa_{1}}(\tilde\eta,\epsilon_c) = 
\frac{U \tilde\eta}
{\Delta E_{{\kappa^{\prime}_{2}\kappa_{2}\kappa^{\prime}_{1}\kappa_{1}}} - \epsilon_c} \ .
\label{etaint}
\end{eqnarray}
Here $\tilde\eta = [1 - \eta (1 - 2 \langle n_{i\uparrow} \rangle_{0}) 
(1 - 2 \langle n_{i\downarrow} \rangle_{0})]$.

The ground-state correlation energy is obtained by substituting the variational 
parameters (\ref{etaint}) into eq. (\ref{ec}).  Each element in the energy is given as follows.
\begin{eqnarray}
\langle \tilde{H} \tilde{O}_{i}\rangle_{0} =
\langle \tilde{O}^{\dagger}_{i}\tilde{H} \rangle_{0}^{\ast} = A\, U^2\, \tilde{\eta} \ ,
\label{ho2}
\end{eqnarray}
\begin{eqnarray}
\langle \tilde{O}^{\dagger}_{i} \tilde{H} \tilde{O}_{i} \rangle_{0} 
= B\, U^2 \tilde \eta^{2}
=\langle \tilde{O}^{\dagger}_{i} \tilde{H}_{0} \tilde{O}_{i} \rangle_{0} 
+ U \langle \tilde{O}^{\dagger}_{i} O_{i} \tilde{O}_{i} \rangle_{0} \ ,
\label{oho2}
\end{eqnarray}
\begin{eqnarray}
\langle \tilde{O}^{\dagger}_{i} \tilde{H}_{0} \tilde{O}_{i} \rangle_{0} 
 =  B_1\, U^2 \tilde \eta^{2} \ ,
\label{oho02}
\end{eqnarray}
\begin{align}
\langle \tilde{O}^{\dagger}_{i} O_{i} \tilde{O}_{i} \rangle_{0} 
= B_2 \, U^2 \, \tilde \eta^2 \ ,
\label{ooo02}
\end{align}
\begin{align} 
\langle \tilde{O}^{\dagger}_{i}\tilde{O}_{i} \rangle_{0} 
=  C\, U^2 \tilde \eta^{2}  \ .
\label{oo2}
\end{align}
Here
\begin{eqnarray}
A=  \int \dfrac{ \Big[\prod\limits^4_{n=1} d\epsilon_{n}\Big]
\rho_{\uparrow}(\epsilon_{1})
\rho_{\uparrow}(\epsilon_{2})\rho_{\downarrow}(\epsilon_{3})
\rho_{\downarrow}(\epsilon_{4})
f(\epsilon_{1})(1-f(\epsilon_{2}))f(\epsilon_{3})(1-f(\epsilon_{4}))
}
{ 
\epsilon_{4} - \epsilon_{3} + \epsilon_{2} - \epsilon_{1} -
\epsilon_{\rm c}
} \ ,
\hspace{3mm}
\label{aetatil}
\end{eqnarray}
\begin{eqnarray}
B=B_1+U\,B_2 \ ,
\hspace{107mm}
\label{betatil}
\end{eqnarray}
\begin{eqnarray}
B_1=  \int \dfrac {\Big[\prod\limits^4_{n=1} d\epsilon_{n}\Big]
\rho_{\uparrow}(\epsilon_{1})
\rho_{\uparrow}(\epsilon_{2})\rho_{\downarrow}(\epsilon_{3})
\rho_{\downarrow}(\epsilon_{4})  
f(\epsilon_{1})(1-f(\epsilon_{2}))f(\epsilon_{3})(1-f(\epsilon_{4}))}
{(\epsilon_{4} - \epsilon_{3} + \epsilon_{2} - \epsilon_{1})^{-1}
(\epsilon_{4} - \epsilon_{3} + \epsilon_{2} - \epsilon_{1} -
\epsilon_{\rm c})^{2}}
 \ ,
\hspace{3mm}
\label{b1etatil}
\end{eqnarray}
\begin{eqnarray}
B_2 &= & \int \frac{\Big[\prod\limits^4_{n=1} d\epsilon_{n}\Big]
\rho_{\uparrow}(\epsilon_{1})
\rho_{\uparrow}(\epsilon_{2})\rho_{\downarrow}(\epsilon_{3})
\rho_{\downarrow}(\epsilon_{4})
f(\epsilon_{1})(1-f(\epsilon_{2}))f(\epsilon_{3})(1-f(\epsilon_{4}))}
{\epsilon_{4} - \epsilon_{3} + \epsilon_{2} - \epsilon_{1} -
\epsilon_{\rm c}}  \hspace{10mm} \nonumber \\
& & \hspace{00mm} \times \bigg[
\int \dfrac{d\epsilon_{5}d\epsilon_{6}
\rho_{\uparrow}(\epsilon_{5})\rho_{\downarrow}(\epsilon_{6})
f(\epsilon_{5})f(\epsilon_{6})}
{\epsilon_{4} - \epsilon_{6} + \epsilon_{2} - \epsilon_{5} -
\epsilon_{\rm c}}
- \int \dfrac{d\epsilon_{5}d\epsilon_{6}
\rho_{\uparrow}(\epsilon_{5})\rho_{\downarrow}(\epsilon_{6})
f(\epsilon_{5})(1-f(\epsilon_{6}))}
{\epsilon_{6} - \epsilon_{3} + \epsilon_{2} - \epsilon_{5} -
\epsilon_{\rm c}}  \nonumber \\
& & \hspace{-15mm} 
- \int \dfrac{d\epsilon_{5}d\epsilon_{6}
\rho_{\uparrow}(\epsilon_{5})\rho_{\downarrow}(\epsilon_{6})
(1-f(\epsilon_{5}))f(\epsilon_{6})}
{\epsilon_{4} - \epsilon_{6} + \epsilon_{5} - \epsilon_{1} -
\epsilon_{\rm c}}
+ \!\! \int \dfrac{d\epsilon_{5}d\epsilon_{6}
\rho_{\uparrow}(\epsilon_{5})\rho_{\downarrow}(\epsilon_{6})
(1-f(\epsilon_{5}))(1-f(\epsilon_{6}))}
{\epsilon_{6} - \epsilon_{3} + \epsilon_{5} - \epsilon_{1} -
\epsilon_{\rm c}}
\bigg] , 
\nonumber \\
\label{b2etatil}
\end{eqnarray}
\begin{eqnarray}
C = \int \dfrac{\Big[\prod\limits^4_{n=1} d\epsilon_{n}\Big]
\rho_{\uparrow}(\epsilon_{1})
\rho_{\uparrow}(\epsilon_{2})\rho_{\downarrow}(\epsilon_{3})
\rho_{\downarrow}(\epsilon_{4})
f(\epsilon_{1})(1-f(\epsilon_{2}))f(\epsilon_{3})(1-f(\epsilon_{4}))}
{(\epsilon_{4} - \epsilon_{3} + \epsilon_{2} - \epsilon_{1} -
\epsilon_{\rm c})^{2}} \ . \hspace{00mm}
\label{cetatil}
\end{eqnarray}
Here $\rho_{\sigma}(\epsilon)$ is the local DOS for the one-electron 
energy eigen values of the HB Hamiltonian matrix (\ref{hbm}).

The best value of $\tilde\eta$ should be determined variationally. 
Infact, when we adopt the approximate form (\ref{etaint}) as a trial 
set of amplitudes, we have a following  inequality 
\begin{equation}
E \leq \langle H \rangle ( w, \{ \eta_{\kappa^{\prime}_{2}\kappa_{2}\kappa^{\prime}_{1}\kappa_{1}}^{\ast}\})
\leq \langle H \rangle (w, \{ \eta_{\kappa^{\prime}_{2}\kappa_{2}\kappa^{\prime}_{1}\kappa_{1} } (\tilde\eta , \epsilon_c)\}) \ .
\end{equation}
Here 
$\{ \eta_{\kappa^{\prime}_{2}\kappa_{2}\kappa^{\prime}_{1}\kappa_{1}}^{\ast}\}$
are the exact solution for the eq. (\ref{eqeta}). The above relation implies that 
the best value of $\tilde{\eta}$  is again determined from the stationary 
condition ($i.e.$,  $\delta \epsilon_c =0$), so that we obtain
\begin{eqnarray}
\tilde{\eta}= \frac{-B + \sqrt{B^2+4A^2CU^2}}{2ACU^2} \ .
\label{betil}
\end{eqnarray}

The total energy per atom should be  obtained 
by taking the configurational average.
\begin{eqnarray}
\langle H \rangle=\overline{ \langle H \rangle}_{\rm HB}+ \overline{\epsilon_c}\ .
\label{tote}
\end{eqnarray}
The HB contribution $\overline{ \langle H \rangle}_{\rm HB}$  has been given 
by eq. $(\ref{ehb})$. The correlation energy can be obtained as follows.
\begin{eqnarray}
\overline{\epsilon_c}=\sum_{\alpha} P_\alpha \,  {\epsilon_{c\alpha}} \,  .
\label{ecbar}
\end{eqnarray} 
Here $\epsilon_{c\alpha}$ denotes the correlation energy for a given on-site
configuration $\alpha$.
\begin{eqnarray}
 {\epsilon_{c\alpha}} = \Big[\dfrac{-\langle
 \tilde{O}^{\dagger}_{i}\tilde{H}\rangle_{0} -
\langle \tilde{H} \tilde{O}_{i} \rangle_{0} + 
\langle \tilde{O}^{\dagger}_{i}\tilde{H}\tilde{O}_{i}\rangle_{0}}
{1 + \langle \tilde{O}^{\dagger}_{i}\tilde{O}_{i} \rangle_{0}}\Big ]_\alpha \ .
\label{eca}
\end{eqnarray}
The quantities 
$\langle \tilde{H} \tilde{O}_{i}\rangle_{0}$, 
$\langle \tilde{O}^{\dagger}_{i} \tilde{H} \tilde{O}_{i} \rangle_{0}$, and 
$\langle \tilde{O}^{\dagger}_{i}\tilde{O}_{i} \rangle_{0}$ are given by eqs.
$(\ref{ho2})$, $(\ref{oho2})$, and $(\ref{oo2})$, respectively,
in which the local DOS have been replaced by those of the single-site
CPA, $i.e.$, eq. (\ref{doshb}).

The double occupation number $\langle n_{i\uparrow}n_{i\downarrow} \rangle$ 
is obtained from $\partial \langle H \rangle / \partial U_{i}$. 
Making use of the single-site energy (\ref{ec}), the 
Feynman-Hellmann theorem~\cite{hell} and taking the 
configurational average, we obtain the following expression.
\begin{eqnarray}
\langle n_{i\uparrow}n_{i\downarrow} \rangle = 
\overline{\langle n_{i\uparrow} \rangle_{0} \langle n_{i\downarrow} \rangle}_{0} 
+ \overline{\langle n_{i\uparrow}n_{i\downarrow} \rangle}_{\rm c} \, ,
%\label{dble}
\end{eqnarray}
Here the HB contribution of the double occupancy
$\overline{\langle n_{i\uparrow} \rangle_{0} 
\langle n_{i\downarrow} \rangle}_{0}$ has been given by eq. $(\ref{dblhb})$.
The second term is the correlation contribution given as follows.
\begin{eqnarray}
\overline{\langle n_{i\uparrow}n_{i\downarrow} \rangle}_{\rm c}=\sum_{\alpha} 
P_\alpha \langle n_{i\uparrow}n_{i\downarrow} \rangle_{\rm{c}\alpha} \, ,
%\label{dble}
\end{eqnarray}
\begin{eqnarray}
\langle n_{i\uparrow}n_{i\downarrow} \rangle_{\rm{c}\alpha}=
\Bigg[\dfrac{-\langle \tilde{O}^{\dagger}_{i} O_{i} \rangle_{0} 
- \langle O_{i} \tilde{O}_{i} \rangle_{0} 
+ \langle \tilde{O}^{\dagger}_{i} O_{i} \tilde{O}_{i} \rangle_{0}
+ \sum_{\sigma} \langle n_{i-\sigma} \rangle_{0} 
\langle \tilde{O}^{\dagger}_{i} \tilde{n}_{i\sigma} \tilde{O}_{i} \rangle_{0}
}
{1+\langle \tilde{O}^{\dagger}_{i} \tilde{O}_{i} \rangle_{0}}\Bigg]_\alpha \, , 
\label{dble}
\end{eqnarray}
\begin{eqnarray}
\langle \tilde{O}^{\dagger}_{i} O_{i} \rangle_{0} 
+ \langle O_{i} \tilde{O}_{i} \rangle_{0} & = &
2 U \tilde{\eta} 
\int \Bigg[\prod\limits^4_{n=1} d\epsilon_{n}\Bigg]
\rho_{\uparrow}(\epsilon_{1})
\rho_{\uparrow}(\epsilon_{2})\rho_{\downarrow}(\epsilon_{3})
\rho_{\downarrow}(\epsilon_{4})\nonumber \\
& & \hspace{10mm}
\times 
\dfrac{f(\epsilon_{1})(1-f(\epsilon_{2}))f(\epsilon_{3})(1-f(\epsilon_{4}))}
{\epsilon_{4} - \epsilon_{3} + \epsilon_{2} - \epsilon_{1} -
\epsilon_{\rm c}} \  ,
\label{otildeo2}
\end{eqnarray}
\begin{eqnarray}
\langle \tilde{O}^{\dagger}_{i} \tilde{n}_{i\sigma} \tilde{O}_{i} 
\rangle_{0} & = & 
U^{2} \tilde{\eta}^2
\int \Bigg[\prod\limits^5_{n=1} d\epsilon_{n}\Bigg]
\rho_{-\sigma}(\epsilon_{1})
\rho_{-\sigma}(\epsilon_{2})\rho_{\sigma}(\epsilon_{3})
\rho_{\sigma}(\epsilon_{4})\rho_{\sigma}(\epsilon_{5})\nonumber \\
& & \hspace{10mm} \times 
\dfrac{f(\epsilon_{1})(1-f(\epsilon_{2}))f(\epsilon_{3})(1-f(\epsilon_{4}))}
{\epsilon_{4} - \epsilon_{3} + \epsilon_{2} - \epsilon_{1} -
\epsilon_{\rm c}} \nonumber \\
& & \hspace{10mm} 
\times \bigg[
\dfrac{1-f(\epsilon_{5})}
{\epsilon_{5} - \epsilon_{3} + \epsilon_{2} - \epsilon_{1} -
\epsilon_{\rm c}}
- \dfrac{f(\epsilon_{5})}
{\epsilon_{4} - \epsilon_{5} + \epsilon_{2} - \epsilon_{1} -
\epsilon_{\rm c}}
\bigg] \, .
\label{ono2}
\end{eqnarray}
The quantities 
$\langle \tilde{O}^{\dagger}_{i} O_{i} \tilde{O}_{i} \rangle_{0}$ and 
$\langle \tilde{O}^{\dagger}_{i}\tilde{O}_{i} \rangle_{0}$ are 
defined by eqs. $(\ref{ooo02})$ and $(\ref{oo2})$, respectively.

Similarly, the momentum distribution $\langle n_{k\sigma} \rangle$ is 
obtained from  $\partial \langle H \rangle / \partial \epsilon_k$
as follows.
\begin{eqnarray}
\langle n_{k\sigma} \rangle = \overline{\langle n_{k\sigma} \rangle}_{0} 
+ \overline{\langle n_{k\sigma} \rangle}_{c}\ .
%\label{nk}
\end{eqnarray}
The HB contribution of the momentum distribution 
$\overline{\langle n_{k\sigma} \rangle}_{0} $ has been given by eq. $(\ref{nkhb})$.
The correlation contribution $\overline{\langle n_{k\sigma} \rangle}_{c}$
is expressed as follows.
\begin{eqnarray}
\overline{\langle n_{k\sigma} \rangle}_{c}=\sum_{\alpha} P_\alpha 
\langle n_{k\sigma} \rangle_{\rm {c}\alpha} \ ,
%\label{nk}
\end{eqnarray}
\begin{eqnarray}
\langle n_{k\sigma} \rangle_{\rm {c}\alpha}=
\Bigg[\dfrac{N \langle \tilde{O}_{i} \tilde{n}_{k\sigma} \tilde{O}_{i} \rangle_{0}}
{1+\langle \tilde{O}^{\dagger}_{i} \tilde{O}_{i} \rangle_{0}}\Bigg]_\alpha \ ,
\label{nk}
\end{eqnarray}
\begin{eqnarray}
N \langle \tilde{O}^{\dagger}_{i} \tilde{n}_{k\sigma} \tilde{O}_{i} 
\rangle_{0} & = &
U^{2} \tilde{\eta}^2 
\int 
 \Bigg[ \prod \limits^4_{n=1} d\epsilon_n\Bigg]
\rho_{\sigma}(\epsilon_{1}) \rho_{-\sigma}(\epsilon_2)\rho_{-\sigma}(\epsilon_{3})
\rho_{k\sigma}(\epsilon_{4}) f(\epsilon_{2})
(1-f(\epsilon_{3})) \nonumber \\
&&
\times \bigg\{
\frac{f(\epsilon_{1})
(1-f(\epsilon_{4}))}
{{(\epsilon_{3} - \epsilon_{2} + \epsilon_{4} - \epsilon_{1} -
\epsilon_{\rm c})^{2}}} 
-
\frac{(1-f(\epsilon_{1}))f(\epsilon_{4})}
{{(\epsilon_{3} - \epsilon_{2} + \epsilon_{1} - \epsilon_{4} -
\epsilon_{\rm c})^{2}}} \bigg\}\ .
\label{onko2}
\end{eqnarray}
Here $\tilde{n}_{k\sigma} = n_{k\sigma} -  \langle n_{k\sigma} \rangle_{0}$.
The DOS in the momentum representation $\rho_{k\sigma}(\epsilon)$ 
has been given by eq. (\ref{mdos}) in the SSA.
The correlation contribution quantity 
$\langle \tilde{O}^{\dagger}_{i}\tilde{O}_{i} \rangle_{0} $ is given by 
eq. $(\ref{oo2})$.

The expressions of these physical quantities
are given by the multiple integrals up to the 6-folds.  One can reduce
these integrals up to the 2-folds using the Laplace transform~\cite{schwe91}.  Their
expressions are given in Appendix B.

In summary, we calculate the correlation energy $\epsilon_{c\alpha}$ 
(eq. (\ref{eca})) self-consistently with use of eqs. (\ref{ho2}), (\ref{oho2}), 
(\ref{oo2}), and  (\ref{betil}) for a given weight $w$, and calculate the 
average correlation energy $\overline{\epsilon_c}$ (eq. (\ref{ecbar})) as well as the 
average HB energy $\overline{\langle H \rangle}_{\rm HB}$  (eq. (\ref{ehb})).
Then we obtain the total energy $\langle H \rangle (w)$ (eq. (\ref{tote}))
for a given $w$. Varying $w$ from $0$ to $1$ numerically, we obtain the 
ground-state energy $\langle H \rangle$. We call this scheme the MLA-HB,
while the simplified scheme in \S 2 the HB. 

\section{Numerical Results: Half-filled band Hubbard Model}
We have performed the  numerical calculations to investigate the 
validity of momentum dependent local-ansatz approach (MLA) with hybrid (HB)
variational wavefunction ($i. e.$,  MLA-HB). We adopted here the 
half-filled band Hubbard model on the  hypercubic lattice in infinite 
dimensions, where the SSA works best~\cite{metz89,kakeh04}, and 
considered  the non-magnetic case.  
In this case, the density of states (DOS) for non-interacting system is given 
by $\rho(\epsilon) = (1/\sqrt{\pi}) \exp (-\epsilon^{2})$~\cite{metz89}. 
The energy unit is chosen to be $\int d\epsilon \rho (\epsilon)\epsilon^{2} = 1/2$.
The characteristic band width $W$ is given by $W=2$ in this unit.

%---------------------------------------------------------------------
\begin{figure}[htbp]
\begin{center}
\includegraphics[scale=0.5,angle=-90]{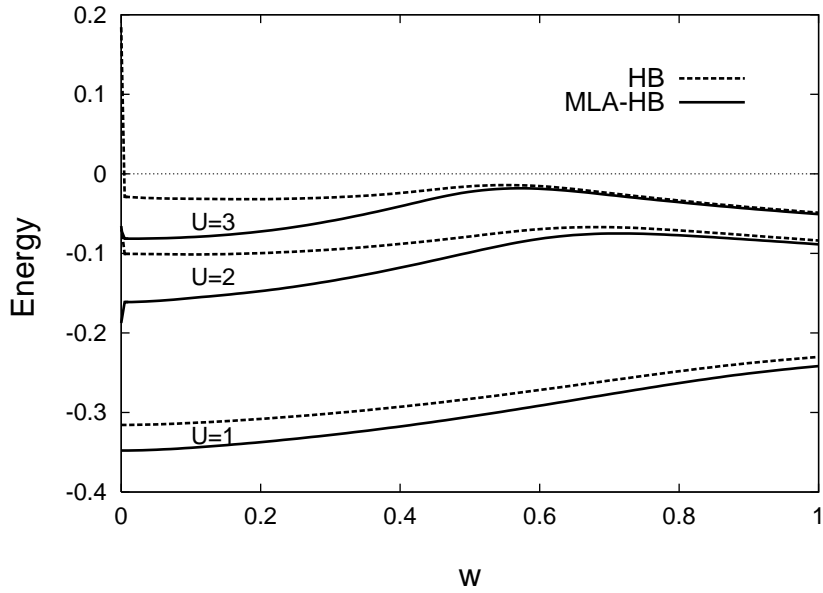}
\caption{\label{wene} Energies as a function of variational parameter $w$
for various Coulomb interaction  parameters $U=1$, $2$, and $3$. 
The HB: dashed curves and the MLA-HB: solid curves.}
\end{center}
\end{figure}
%--------------------------------------------------------------------- 
We  calculated the energy for given variational parameter $w$
by minimizing it with respect to the variational parameter $\tilde\eta$ 
self-consistently. Figure \ref {wene} shows the calculated energy vs $w$ curves 
for various Coulomb interactions, $U=1$, $2$, and $3$. In the HB scheme  
(without correlator) for $U=1$, the energy increases 
monotonically with increasing $w$. Therefore the Fermi liquid HF state ($w=0$) 
is stabilized as the ground state.  When we increase the $U$ value, the  
HF energy continues to increase, while the random-potential states with  
$w\neq 0$, which is driven by the AA potential, are relatively stabilized. 
Such a random-potential state remains even at $w=\varepsilon$ 
as shown in the dashed curve $U=2$, where  $\varepsilon$ is the 
infinitesimal positive number.
It is caused by the HF-type self-consistent random potentials and is accompanied
by the disordered local moments~\cite{kakeh80,dgpet80}. 
When we further increase $U$, the AA state ($w=1$) is more stabilized 
(see the dashed curve $U=3$), so that we find the first-order transition at $U=2.31$, 
and the insulating state is realized. We also find the similar behavior for 
the MLA-HB, in which the transition takes place at $U=2.81$. 
However it should be 
noted that the transition occurs between the Fermi liquid state ($w=0$) and the 
disordered local moment state ($w=\epsilon$) in the case of the MLA-HB;
we found numerically that the latter  ($w=\epsilon$) remains stable as
compared with the AA state ($w=1$) even if we increase further the Coulomb 
interaction strength $U$.
 
%---------------------------------------------------------------------
\begin{figure}[htbp]
\begin{center}
\includegraphics[scale=0.5,angle=-90]{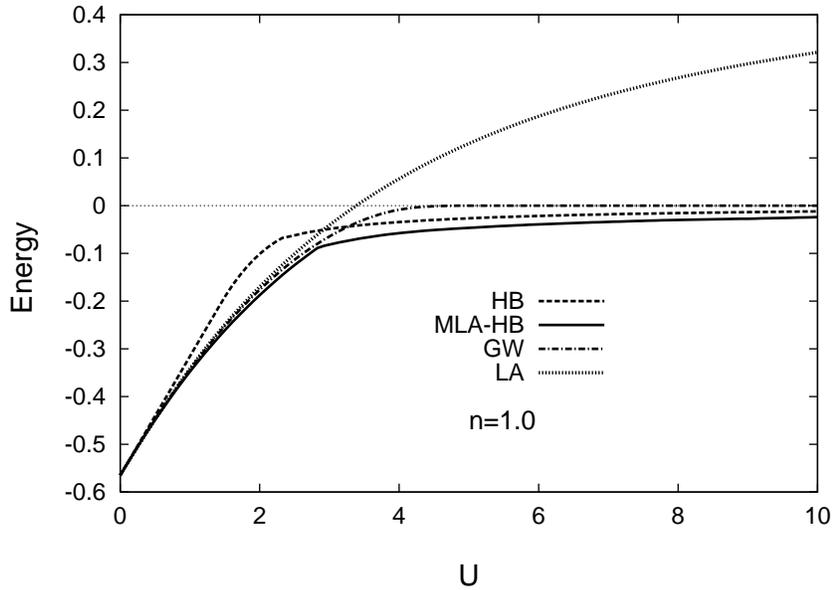}
\caption{\label{energy} The energy vs Coulomb interaction energy
$U$ curves in the HB (dashed curve), the MLA-HB (solid curve), 
the GW (dot-dashed curve) and the LA (dotted curve) for the electron number $n=1.0$.}
\end{center}
\end{figure}
%---------------------------------------------------------------------
The results of the ground-state  energy vs Coulomb 
interaction energy curves are shown in Fig. \ref{energy}. 
The energy of the HB wavefunction linearly
increases with increasing Coulomb interaction strength $U$ in the 
weak $U$ regime. At $U_{c0}=1.43$, the system shows a transition 
from the Fermi liquid (FL) state ($w=0$) to a non-Fermi liquid (NFL)
state ($w\neq 0$), and shows a kink at the critical Coulomb 
interaction $U_{c}=2.31$, indicating the metal-insulator transition.
The transition is of the first order in the present approach. 
The HB  wavefunction  gives lower energy 
in comparison with the GW  and the LA in the strong Coulomb interaction 
regime ($U/W\gtrsim 1.5$). The MLA-HB wavefunction further lowers the 
energy. In the weak Coulomb interaction regime, the ground-state 
energy of the MLA-HB is the lowest among the HB, LA, GW, and the
MLA-HB. The MLA-HB shows the first-order transition at $U_{c}=2.81$
from the FL to the NFL, indicating the metal-insulator transition. 
The MLA-HB scheme gives lower energy for overall 
Coulomb  interaction and therefore overcomes the GW.

%---------------------------------------------------------------------
\begin{figure}[htbp]
\begin{center}
\includegraphics[scale=0.5, angle=-90]{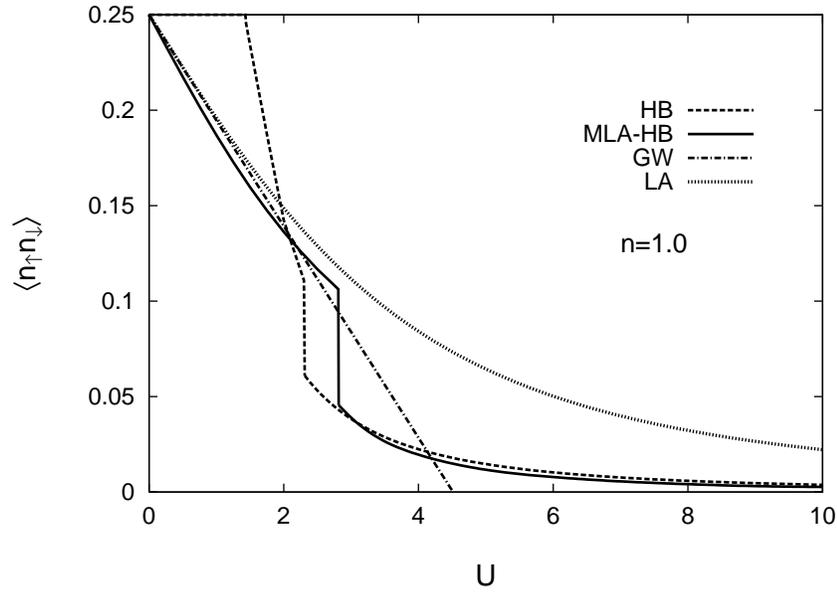}%
\caption{\label{dbl}The double occupation number  $\langle n_{\uparrow}n_{\downarrow} \rangle$ 
vs Coulomb interaction energy $U$ curves at half-filling ($n=1.0$) in 
the HB (dashed curve), the MLA-HB (solid curve), 
the GW (dot-dashed curve), and the LA (dotted curve). }
\end{center}
\end{figure}
%---------------------------------------------------------------------
%
Figure \ref{dbl} shows  the double occupation number
$\langle n_{\uparrow}n_{\downarrow} \rangle$ as a function of  Coulomb 
interaction energy $U$ at half-filling. In the case of  the HB,  
the double occupancy is constant ($1/4$) up to $U_{c0}=1.43$, and 
decreases rapidly  up to the critical point $U_{c}=2.31$, at which 
it jumps from $0.110$ to $0.060$.
In the strong Coulomb interaction  regime  the double occupancy  
decreases with increasing $U$ and vanishes in the atomic limit.  
In the case of the MLA-HB, the double occupation number decreases  
smoothly from $1/4$ with increasing Coulomb interaction so as  to reduce 
the loss of Coulomb energy $U$. Note that the MLA-HB reduces more the 
double occupancy  as compared with  that of the HB, GW and the LA  in the 
weak $U$ region. The double occupancy in the MLA-HB jumps from 
$0.106$ to $0.045$ at the transition point $U_{c}=2.81$, and again 
monotonically decreases with increasing $U$. 
Note that the double occupancy in the MLA-HB remains finite
in the strong $U$ regime  as it should be, 
while the GW gives the Brinkman-Rice atom. 

%---------------------------------------------------------------------
\begin{figure}[htbp]
\begin{center}
\includegraphics[scale=0.5, angle=-90]{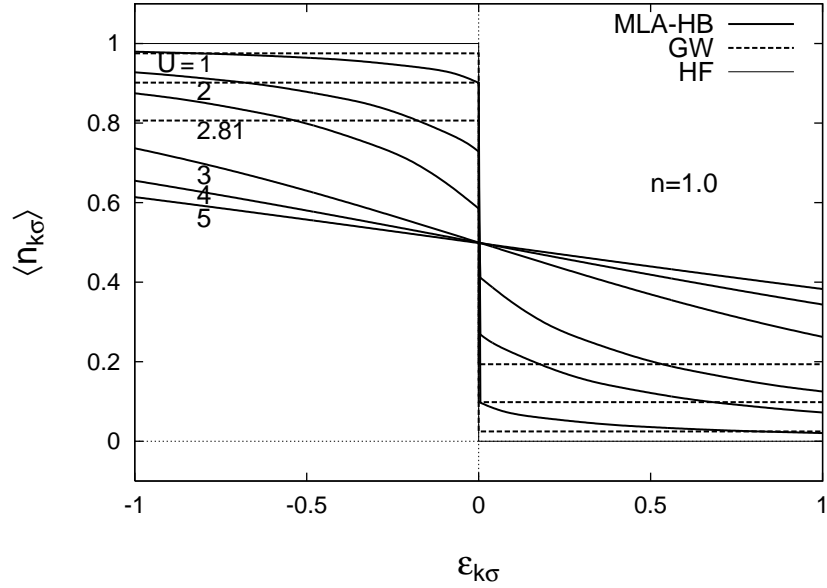}%
\caption{\label{momentum} The momentum distribution as a function of
energy  $\epsilon_{k\sigma}$ for various Coulomb interaction energy parameters
$U=1.0,2.0,2.81,3.0,4.0$ and $5.0$ at half-filling. The MLA-HB: solid curves, 
the GW: dashed curves, and the HF: thin solid curve.}
\end{center}
\end{figure}
%---------------------------------------------------------------------
The momentum distribution for the MLA-HB is shown in Fig. \ref{momentum}.
It decreases monotonically with increasing $\epsilon_{k\sigma}$ 
$(=\epsilon_0-\mu+\epsilon_k)$ and shows a 
jump at the Fermi energy in the metallic regime. The jump decreases with 
increasing $U$, and disappears beyond $U_{c}$. When we  further increase the 
Coulomb interaction  $U$ the curve becomes flatter. Note that  the 
momentum distributions for the GW are constant below and above the Fermi 
level~\cite{gutz63,gutz64,gutz65}. These results indicate that the MLA-HB
improves the GW. 

%---------------------------------------------------------------------
\begin{figure}[htbp]
\begin{center}
\includegraphics[scale=0.5, angle=-90]{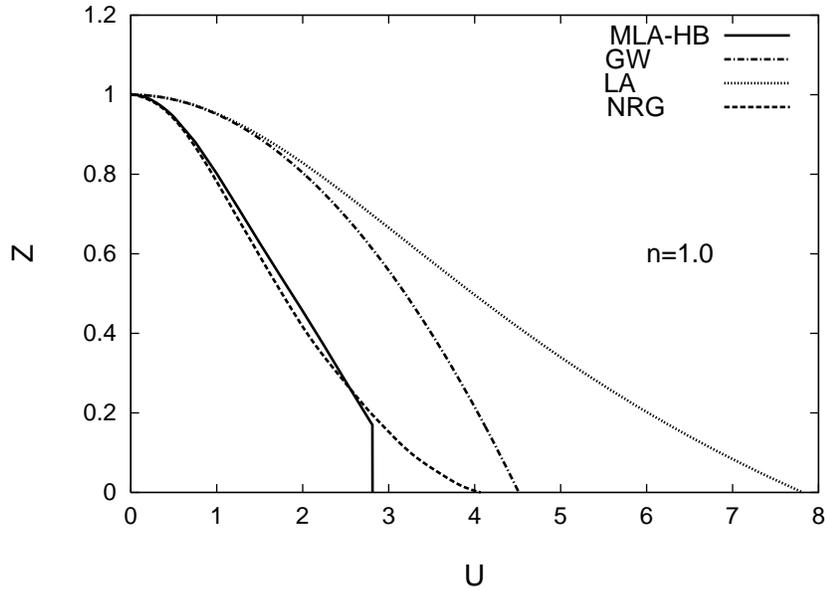}%
\caption{\label{quasiparticlew}
Quasiparticle-weight vs. Coulomb interaction curves in various theories.
The MLA-HB: solid curve, the GW: dot-dashed, the LA: dotted curve, and 
the NRG: dashed curve~\cite{rbulla99}. }
\end{center}
\end{figure}
%---------------------------------------------------------------------
The quasiparticle weight $Z$ ($i.e.$, inverse effective mass) is obtained 
from the jump at the Fermi level  in the momentum distribution according 
to the Fermi liquid  theory~\cite{landu57,gbaym91}. Calculated 
quasiparticle weight vs Coulomb interaction curves are shown in 
Fig. \ref{quasiparticlew}. The GW and the LA  
curves strongly deviate from the curve of the NRG~\cite{rbulla99} which 
is considered to be the best. The MLA-HB is close to the NRG in the 
metallic regime, and vanishes beyond $U_c=2.81$. It should be noted 
that the NRG~\cite{rbulla99} also shows the first-order transition at a critical 
Coulomb interaction $U_c$ before $Z$ vanishes at $U_{c2}=4.1$.
The values of $U_c$ in the NRG, however, has not yet been published.

\section{Summary and Discussions}
We have proposed a new hybrid (HB) wavefunction and combined it with 
the momentum dependent local-ansatz approach MLA ($i.e.,$ the MLA-HB) to describe
the correlated electron system from the weak to the strong Coulomb 
interaction regime. The HB wavefunction as a starting wavefunction 
is the ground-state for the HB Hamiltonian. The latter was constructed 
as a superposition of the Hartree-Fock (HF) Hamiltonian and the 
alloy-analogy (AA) one. The weight $w$ of superposition is regarded as 
a variational parameter. When we adopt  $w=0\,(1)$, the HB wavefunction
reduces to the HF (AA) state. In the MLA-HB, the best wavefunction 
is chosen by controlling the momentum dependent variational parameters 
for  the two-particle excited states as well as the HB parameter $w$. 
We obtained the ground-state energy of the MLA-HB within a single-site 
approximation, and  derived an approximate solution  for the self-consistent 
equations of  the variational  parameters which interpolates between the 
weak Coulomb interaction limit  and the atomic limit. 

To examine the improvement and validity of the theory,
we have performed the  numerical calculations for 
the half-filled band Hubbard model  on the  hypercubic lattice in infinite 
dimensions. In case of the HB wavefunction we clarified that 
the ground-state energy increases linearly in the weak $U$
regime and it shows a lower energy as compared with the GW and the LA in the strong 
$U$ regime. The double occupation number is constant  up to 
the  $U=1.43$ ($i.e.,$ $\langle n_{\uparrow}n_{\downarrow} \rangle_{\rm HB}=0.25$) 
and then decreases rapidly to the critical value $U_c=2.31$ 
where the first-order metal-insulator transition occurs. In the strong $U$ regime the 
$\langle n_{\uparrow}n_{\downarrow} \rangle_{\rm HB}$  remains finite.

We have demonstrated that the ground-state energy of the MLA-HB is  lower 
than that of  the  HB, GW and the LA  in the whole Coulomb interaction regime.
In the weak and intermediate Coulomb interaction regimes, the double occupation 
number is suppressed as compared with the others. It jumps at $U_{c}=2.81$ 
and remains finite in the strongly correlated regime as it 
should be. The momentum distribution functions show  a distinct  
momentum dependence in both the weak and the strong $U$ regimes.
Moreover, we found that the behavior of the quasiparticle weight is 
close to the NRG one. The above mentioned results 
indicate that the MLA-HB approach overcomes the limitations of the original MLA~\cite{kakeh08, pat11},
and describes reasonably correlated electrons from the weak to the  
strong Coulomb interaction regime, so that it goes beyond the GW in 
the whole Coulomb interaction  $U$ regime. 
Although advanced theories based on the QMC and the NRG have been developed,
the MLA-HB approach presented in this work   is applicable to more complex 
systems and allows us to calculate any static averages with use of the 
wavefunction. Further developments of the MLA wavefunction  approach 
should provide us  with a useful tool for understanding the properties
of correlated electrons and their physics in  the realistic systems.

\section*{Acknowledgments}

The present work is supported by a Grant-in-Aid for Scientific Research
(22540395) in the MEXT (Japan). 
%============================================================================
\appendix
\section{Derivation of $P_2$}
In the derivation of the hybrid (HB) Hamiltonian we made the following approximations for the 
alloy-analogy (AA) and the Hartree-Fock (HF) Hamiltonians, respectively.
\begin{eqnarray}
\hat n_\uparrow \hat n_\downarrow \approx n_\uparrow  \hat n_\downarrow
+n_\downarrow  \hat n_\uparrow 
-n_\uparrow n_\downarrow\, \, \, \, \, \, (\rm AA) \ ,
\end{eqnarray}
\begin{eqnarray}
\hat n_\uparrow \hat n_\downarrow \approx \hat n_\uparrow \langle \hat n_\downarrow\rangle
+\hat n_\downarrow \langle \hat n_\uparrow \rangle
-\langle \hat n_\uparrow \rangle \langle \hat n_\downarrow \rangle  \, \, \, \, \, \, (\rm HF)\ .
\end{eqnarray}
In the HB scheme, we approximate the averages $\langle \sim \rangle$ at the rhs of the above 
expressions with those of the HB Hamiltonian (\ref{hb}), and superpose 
them with the weight $w$ and ($1-w$), respectively. Taking the 
the quantum mechanical average of the superposed double occupation number 
as well as the configurational average, we obtain the probability of the double occupation
$P_2$ $(=\overline {\langle n_\uparrow  n_\downarrow\rangle})$ in the
HB approximation as follows.
\begin{eqnarray}
P_2 &=& w \, ( \overline {n_\uparrow \langle n_\downarrow\rangle}
+\overline {n_\downarrow \langle  n_\uparrow \rangle}
-\overline {n_\uparrow n_\downarrow})
+ (1-w) \, \overline {\langle n_\downarrow \rangle \langle  n_\uparrow \rangle}\nonumber \\
&=& w \, \overline {n_\uparrow \langle n_\downarrow\rangle}
+(1-w) \, \overline {\langle  n_\uparrow \rangle \langle n_\downarrow \rangle }
+  w\, \overline {n_\downarrow \langle  n_\uparrow \rangle}
+ (1-w) \, \overline {\langle  n_\uparrow \rangle \langle n_\downarrow \rangle }\nonumber \\
& & \hspace{5cm}
-[w\, \overline {n_\uparrow n_\downarrow}+(1-w) \, 
\overline {\langle  n_\uparrow \rangle \langle n_\downarrow \rangle }] \ .
\label{app1}
\end{eqnarray}
The last term at the rhs of eq. (\ref{app1}) may be regarded as the probability
$P_2$. Therefore, we obtain
\begin{eqnarray}
P_2 & \approx & \frac{1}{2}\big [ \overline{  (w \,  n_\uparrow+(1-w)\langle n_\uparrow\rangle ) \langle n_\downarrow\rangle}
+ \overline{  (w \,  n_\downarrow+(1-w)\langle n_\downarrow\rangle ) \langle n_\uparrow\rangle} \big] \nonumber \\
&= & \frac{1}{2}w \, ( \overline {n_\uparrow \langle n_\downarrow\rangle}
+\overline {n_\downarrow \langle  n_\uparrow \rangle})
+(1-w) \, \overline {\langle  n_\uparrow \rangle\langle n_\downarrow \rangle } \ .
\label{app2}
\end{eqnarray}
In the single-site approximation, the last term of the rhs is expressed 
as follows.
\begin{eqnarray}
\overline{\langle n_{ \uparrow} \rangle \langle n_{\downarrow} \rangle} &=&\sum_{\alpha} 
P_\alpha \langle n_{ \uparrow} \rangle_{\alpha} \langle n_{\downarrow}\rangle_{\alpha} \nonumber \\
&=&P_0\langle n_{ \uparrow} \rangle_{00} \langle n_{\downarrow}\rangle_{00}
+P_{1\uparrow}\langle n_{ \uparrow} \rangle_{10} \langle n_{\downarrow}\rangle_{10}
+P_{1\downarrow}\langle n_{ \uparrow} \rangle_{01} \langle n_{\downarrow}\rangle_{01}
+P_2\langle n_{ \uparrow} \rangle_{11} \langle n_{\downarrow}\rangle_{11} \ . \nonumber \\
\end{eqnarray} 
In the non-magnetic case, we have
\begin{eqnarray}
\overline{\langle n_{ \uparrow} \rangle \langle n_{\downarrow} \rangle}
&=& \langle n_{ \uparrow} \rangle^2_{00}+ (P_\uparrow+P_\downarrow)
(\langle n_{ \uparrow} \rangle_{10}\langle n_{ \uparrow} \rangle_{01}-\langle n_{ \uparrow} \rangle^2_{00})\nonumber \\
&&\hspace{30mm}
+P_2(\langle n_{ \uparrow} \rangle^2_{00}-2\langle n_{ \uparrow} \rangle_{10}
\langle n_{ \uparrow} \rangle_{01}+\langle n_{ \uparrow} \rangle^2_{11})\ .
\label{apdblhb}
\end{eqnarray} 
Similarly,
\begin{eqnarray}
\overline{n_{\uparrow}\langle n_{ \downarrow} \rangle}+ \overline{n_{\downarrow}\langle n_{ \uparrow} \rangle}
 \, = \, (P_\uparrow+P_\downarrow)\langle n_{ \uparrow} \rangle_{01}
 + 2 P_2 \, (\langle n_{ \uparrow} \rangle_{11}-\langle n_{i \uparrow} \rangle_{01}) \ . \hspace{30mm}
 \label{apdblhb2}
\end{eqnarray} 
Substituting (\ref{apdblhb}) and  (\ref{apdblhb2}) into eq. (\ref{app2}), we 
obtain the final expression of $P_2$, $i.e.$, eq. (\ref{p2}).
\begin{eqnarray}
P_2=\dfrac{(1-w)\langle n_{\uparrow}\rangle^{2}_{00}+(P_\uparrow+P_\downarrow)
\{1/2\, w \,\langle n_{\uparrow}\rangle_{01} + (1-w)\langle n_{\uparrow}\rangle_{10}
\langle n_{\uparrow}\rangle_{01}- \langle n_{\uparrow}\rangle^{2}_{00}\}}
{1-w(\langle n_{\uparrow}\rangle_{11}-\langle n_{\uparrow}\rangle_{01})
-(1-w)(\langle n_{\uparrow}\rangle^{2}_{00}-2\langle n_{\uparrow}\rangle_{10}
\langle n_{\uparrow}\rangle_{01}+\langle n_{\uparrow}\rangle^{2}_{11} )} \, . 
%\label{p2}
\end{eqnarray}

%============================================================================
\section{Laplace Transform for Correlation Calculations} 
 
The Laplace transform can significantly reduce the number of integrals in 
the physical quantities which appear in our variational theory. It is  written as follows.
\begin{eqnarray}
\dfrac{1}{z - \epsilon_{4} + \epsilon_{3} - \epsilon_{2} + \epsilon_{1}
 + \epsilon_{c}} = -i \int^{\infty}_{0} dt \,
{\rm
e}^{i(z-\epsilon_{4}+\epsilon_{3}-\epsilon_{2}+\epsilon_{1}
+\epsilon_{\rm c}) \, t}
\ .
\label{laplace}
\end{eqnarray}
Here $z=\omega+i\delta$, and $\delta$ is an infinitesimal positive
number. 

Laplace transforms of the physical quantities (\ref{ho2})-(\ref{oo2})
are given as follows.
\begin{eqnarray}
 A_\alpha = i  \int^{\infty}_{0} \! dt \, {\rm e}^{i{ \epsilon_{c\alpha}}t} \, 
a_{\alpha\uparrow}(-t)a_{\alpha\downarrow}(-t)b_{\alpha\uparrow}(t)b_{\alpha\downarrow}(t) \ ,
\label{lho}
\end{eqnarray}
\begin{eqnarray}
B_{1\alpha} & = & - 
\int^{\infty}_{0} \! dtdt^{\prime} 
{ e}^{i{ \epsilon_{c\alpha}}(t+t^{\prime})} \hspace{00mm}\nonumber \\
& & \hspace{00mm}
\times \big[
a_{\alpha\uparrow}(-t-t^{\prime})b_{\alpha\uparrow}(t+t^{\prime})
a_{\alpha\downarrow}(-t-t^{\prime})b_{1\alpha\downarrow}(t+t^{\prime})  \nonumber \\
& & \hspace{00mm}
-a_{\alpha\uparrow}(-t-t^{\prime})b_{\alpha\uparrow}(t+t^{\prime})
a_{1\alpha \downarrow}(-t-t^{\prime})b_{\alpha \downarrow}(t+t^{\prime})  \nonumber \\
& & \hspace{00mm}
+a_{\alpha\uparrow}(-t-t^{\prime})b_{ 1\alpha\uparrow}(t+t^{\prime})
a_{\alpha \downarrow}(-t-t^{\prime})b_{\alpha \downarrow}(t+t^{\prime})  \nonumber \\
& & \hspace{00mm}
-a_{1\alpha \uparrow}(-t-t^{\prime})b_{\alpha \uparrow}(t+t^{\prime})
a_{\alpha \downarrow}(-t-t^{\prime})b_{\alpha \downarrow}(t+t^{\prime})
\big] \ , \hspace{00mm}
\label{loh0o}
\end{eqnarray} 
\begin{eqnarray}
B_{2\alpha} & = &
- 
\int^{\infty}_{0} \! dtdt^{\prime} 
{\rm e}^{i {\epsilon_{c\alpha}}(t+t^{\prime})}\nonumber \\
&& \times
\big[
a_{\alpha\uparrow}(-t)b_{\alpha\uparrow}(t+t^{\prime})
a_{\alpha\downarrow}(-t)b_{\alpha\downarrow}(t+t^{\prime})
a_{\alpha\uparrow}(-t^{\alpha\prime})a_{\alpha\downarrow}(-t^{\prime})  \nonumber \\
& & \hspace{00mm}
-a_{\alpha\uparrow}(-t)b_{\alpha\uparrow}(t+t^{\prime})
a_{\alpha\downarrow}(-t-t^{\prime})b_{\alpha\downarrow}(t)
a_{\alpha\uparrow}(-t^{\prime})b_{\alpha\downarrow}(t^{\prime})  \nonumber \\
& & \hspace{00mm}
-a_{\alpha\uparrow}(-t-t^{\prime})b_{\alpha\uparrow}(t)
a_{\alpha\downarrow}(-t)b_{\alpha\downarrow}(t+t^{\prime})
b_{\alpha\uparrow}(t^{\prime})a_{\alpha\downarrow}(-t^{\prime})  \nonumber \\
& & \hspace{00mm}
+a_{\alpha\uparrow}(-t-t^{\prime})b_{\alpha\uparrow}(t)
a_{\alpha\downarrow}(-t-t^{\prime})b_{\alpha\downarrow}(t)
b_{\alpha\uparrow}(t^{\prime})b_{\alpha\downarrow}(t^{\prime})
\big] \ , \hspace{00mm}
\label{looo}
\end{eqnarray}
\begin{eqnarray}
C_\alpha  = -  \int^{\infty}_{0} \! dtdt^{\prime} 
{\rm e}^{i {\epsilon_{c\alpha}}(t+t^{\prime})}
a_{\alpha\uparrow}(-t-t^{\prime})b_{\alpha\uparrow}(t+t^{\prime})
a_{\alpha\downarrow}(-t-t^{\prime})b_{\alpha\downarrow}(t+t^{\prime}).
\end{eqnarray}
Here $\alpha$ denotes the local electron configuration $(\alpha=0,\, 1\uparrow,\, 1\downarrow, \, 2)$, and 
\begin{eqnarray}
a_{\alpha\sigma}(t) = \int d\epsilon \, \rho^{\alpha}_{\sigma}(\epsilon)
f(\epsilon) \, {\rm e}^{-i\epsilon t}
\ ,
\end{eqnarray}
\begin{eqnarray}
b_{\alpha\sigma}(t) = \int d\epsilon \, \rho^{\alpha}_{\sigma}(\epsilon)
[1-f(\epsilon)] \, {\rm e}^{-i\epsilon t}
\ ,
\end{eqnarray}
\begin{eqnarray}
a_{1\alpha\sigma}(t) = \int d\epsilon \, \rho^{\alpha}_{\sigma}(\epsilon)
f(\epsilon)\, \epsilon \, {\rm e}^{-i\epsilon t}
\ ,
\end{eqnarray}
\begin{eqnarray}
b_{1\alpha\sigma}(t) = \int d\epsilon \, \rho^{\alpha}_{\sigma}(\epsilon)
[1-f(\epsilon)] \, \epsilon \, {\rm e}^{-i\epsilon t}
\ .
\end{eqnarray}

The element (\ref{otildeo2}) for the calculation of the double occupancy
is expressed as
\begin{eqnarray}
\langle \tilde{O}^{\dagger}_{i} O_{i} \rangle_{0\alpha} 
+ \langle O_{i} \tilde{O}_{i} \rangle_{0\alpha} & = &
2iU \tilde \eta_\alpha
\int^{\infty}_{0} \! dt \,
{\rm e}^{i {\epsilon_{c\alpha}} t}
a_{\alpha\uparrow}(-t)b_{\alpha\uparrow}(t)
a_{\alpha\downarrow}(-t)b_{\alpha\downarrow}(t) \ . 
\hspace{15mm}
\label{lotildeo}
\end{eqnarray}
The correlation contribution to the electron number (\ref{ono2}) 
which appears in the
calculation of the double occupation number is expressed as
\begin{eqnarray}
\hspace{-8mm}
\langle \tilde{O}^{\dagger}_{i} \tilde{n}_{i\sigma} \tilde{O}_{i} \rangle_{0\alpha} 
& = & 
- U^{2} \tilde\eta^{2}_{\alpha}   
\int^{\infty}_{0} \! dtdt^{\prime} 
{\rm e}^{i {\epsilon_{c\alpha}}(t+t^{\prime})}\nonumber \\
& & \times
\big[
a_{\alpha(-\sigma)}(-t-t^{\prime})b_{\alpha(-\sigma)}(t+t^{\prime})
a_{\alpha\sigma}(-t-t^{\prime})b_{\alpha\sigma}(t)b_{\alpha\sigma}(t^{\prime}) \nonumber \\
& &  \hspace{00mm}
-a_{\alpha(-\sigma)}(-t-t^{\prime})b_{\alpha(-\sigma)}(t+t^{\prime})
a_{\alpha\sigma}(-t)b_{\alpha\sigma}(t+t^{\prime})a_{\alpha\sigma}(t^{\prime})
\big] \ .  \hspace{00mm}
\label{lonio}
\end{eqnarray}

The correlation contribution to the momentum distribution function
(\ref{onko2}) is given by
\begin{eqnarray}
N \langle \tilde{O}^{\dagger}_{i} \tilde{n}_{k\sigma} \tilde{O}_{i} \rangle_{0\alpha} 
& = & 
U^{2} \tilde \eta^{2}_\alpha 
\int^{\infty}_{0} \! dtdt^{\prime} 
{e}^{i {\epsilon_{c\alpha}}(t+t^{\prime})}
a_{\alpha(-\sigma)}(-t-t^{\prime})b_{\alpha(-\sigma)}(t+t^{\prime}) \nonumber \\
& &  \hspace{00mm}  \times
\big[ b_{\alpha\sigma}(t+t^{\prime}) a_{k\sigma}(-t-t^{\prime}) 
-a_{\alpha\sigma}(-t-t^{\prime})b_{k\sigma}(t+t^{\prime}) 
\big]
\ .
\label{lonko}
\end{eqnarray}
Here
\begin{eqnarray}
a_{k\sigma}(t) = \int d\epsilon \, \rho_{k\sigma}(\epsilon)
f(\epsilon) \, {\rm e}^{-i\epsilon t}
\ ,
\end{eqnarray}
\begin{eqnarray}
b_{k\sigma}(t) = \int d\epsilon \, \rho_{k\sigma}(\epsilon)
 \big[1-f(\epsilon)\big] \, {\rm e}^{-i\epsilon t}
\ .
\end{eqnarray}
\end{document}